\RequirePackage{fix-cm}
\documentclass[conference,onecolumn]{IEEEtran}
\usepackage{graphicx}

\def\sinv{\vspace*{-2.5pt}}
\def\inv{\vspace*{-6pt}}

\usepackage{comment}
\usepackage{verbatim}
\usepackage{amsmath, amssymb}
\usepackage{subfigure}
\usepackage{times}
\usepackage{algorithm}
\usepackage{algorithmic}
\usepackage{epstopdf}
\usepackage{mathrsfs}
\usepackage{url}
\usepackage{multirow}
\usepackage{color}
\usepackage{setspace}
\usepackage{epsfig}

\begin{document}

\title{Secure Personal Content Networking \\over Untrusted Devices%\thanks{Grants or other notes
}


\author{\IEEEauthorblockN{Uichin Lee}
\IEEEauthorblockA{KAIST\\
uclee@kaist.edu}
\and
\IEEEauthorblockN{Joshua Joy}
\IEEEauthorblockA{UCLA\\
jjoy@cs.ucla.edu}
\and
\IEEEauthorblockN{Youngtae Noh}
\IEEEauthorblockA{CISCO\\
ynoh@cisco.com}}


%\institute{YoungTae Noh is the corresponding author.\\
%              Tel.: +1-323-303-7945\\
%              \email{ynoh@cisco.com}           %  \\
%}

\date{Received: date / Accepted: date}

\maketitle

\begin{abstract}
Securely sharing and managing personal content is a challenging task in multi-device environments. In this paper, we design and implement a new platform called Personal Content Networking (PCN). Our work is inspired by Content-Centric Networking (CCN) because we aim to enable access to personal content using its name instead of its location. The unique challenge of PCN is to support secure file operations such as replication, updates, and access control over distributed untrusted devices. The primary contribution of this work is the design and implementation of a secure content management platform that supports secure updates, replications, and fine-grained content-centric access control of files. Furthermore, we demonstrate its feasibility through a prototype implementation on the CCNx skeleton.

%\keywords{Information Centric Networking, Personal Content Sharing, Content-Centric Access Control}
\end{abstract}

\section{Introduction}
Today, people carry various consumer electronic devices such as digital cameras, smartphones, and laptops. These Internet-enabled smart devices enable both \emph{consumers} of published content and \emph{producers} of user-generated content. Content creation has become very easy because anyone can post content using Web 2.0 tools, e.g. YouTube, Flickr, Twitter, etc. As a result, personal content is exploding: content is shared and stored in multiple places ranging from personal devices to cloud storage. A recent report estimated that by 2015, terabytes of data will be in a person's pocket and petabytes of data in a person's home~\cite{Storm07}. Under these circumstances, it is very important to have a system that seamlessly enables networking of personal content such that users can manage personal content across multiple devices (including cloud storage) and selectively share content with intended groups (e.g. family members, friends, and colleagues).

The first step toward this goal is to introduce single persistent naming over personal content across multiple devices. Because the current generation of personal devices maintains individual namespaces in each device, content is closely tied to a device (i.e. it is location dependent). As the amount of content increases, content management becomes more difficult because users tend to lose track of what files are located where. A unified view with persistent naming will allow users to make location-independent (or content-centric) queries where there is no need to specify which device has the requested content. For example, Alice can access her favorite songs via a name: ``\emph{Alice/Music/My Favorites}''. Similarly, she can share the collection to Bob by simply telling him the location name.

This content centric approach is considered to be a key feature of the future Internet through Content-Centric Networking (CCN), which replaces the conventional host-to-host conversations with name-based communications and provides secure binding between the name and data in order to thwart security attacks. The name-based routing of CCN enables content retrieval over a fully distributed network without specifying where the content is located because any nodes that have the requested data locally answer the request~\cite{Jacobson09}. While CCN was originally designed for large-scale content dissemination (or potentially replacing the existing IP network), its key principles (i.e. name-based routing and secure binding) are also applicable in the realization of secure personal content networking. However, due to scalability and performance reasons, CCN transfers data and also caches data on \emph{untrusted devices} that forward/store data properly and yet do not necessarily keep the data confidential. In addition, CCN lacks the essential component of personal content networking of secure content management, such as content updates and access control over untrusted nodes.

While secure content management is an active area of research in the field of distributed file systems, existing work has primarily focused on host-centric trusted file systems; a trusted file server handles user authentication and access control authorization, and then provides data confidentiality through securing the communication channel (e.g. SFS~\cite{Mazieres99}). When managing untrusted storage, the files must be encrypted in order to assure data confidentiality, e.g. Cryptographic File System (CFS)~\cite{Blaze93} and Plutus~\cite{Kallahalla03}. However, such cryptographic storage systems cannot generally provide fine-grained expressive access control; for example, in Plutus, a file can only be encrypted using a single key. If a user wants to share the file with more than two groups, it is not clear which key should be used for the encryption. A simple solution is to use a common key for file encryption and to encrypt this key using each user's public key as in SiRiUS~\cite{Goh03}. However, this approach is limited because the metadata size linearly scales with the number of users, and supporting more expressive access control is difficult because it only uses a single file encryption key. Moreover, the existing cryptographic systems do not support \emph{secure binding} between the name and data and, as a result, the channel must be secured in order to prevent man-in-the-middle attacks.

In this paper, we propose the Personal Content Networking (PCN) platform and it provides a secure content management mechanism over CCN, which enables secure replication and updates, as well as fine-grained content-centric access control. We extend CCN to build a framework for distributed content management with replication and updates. Then, we propose and implement a \emph{secure content-centric access control} mechanism using the recently proposed cryptography tool called Attribute-Based Encryption (ABE), which permits secure sharing of content within a group over untrusted devices~\cite{Bethencourt07}. ABE supports fine-grained expressive access policies called Attribute-Based Access Control (ABAC). An owner can define a set of attributes (e.g. college friends, CS219 team, family members, etc.) and then they issue a secret key for the assigned attributes to an individual. Each file is encrypted based on the access policy over the attributes using the owner's public key. For a given encrypted file and access policy, any user can decrypt the file as long as they have the secret key with the attributes that satisfy the given policy.

While ABE was designed and has been used for selective \emph{read-only} content sharing over untrusted storage~\cite{Yu08,Yu10}, this is the first attempt to build a fully distributed personal storage system that supports ABE-based fine-grained access control with \emph{read-write} operations over untrusted devices and \emph{secure-binding} between the name and data. The primary contribution of this paper is twofold.
\begin{itemize}
\item We design the PCN platform through significantly extending CCN in order to realize a secure content management mechanism that supports secure replication and updates, as well as fine-grained content-centric access control.
\item We build a PCN prototype through integrating the whole system using FUSE~\cite{Fuse}, which is a user level file system, and we demonstrate that a user can seamlessly access and manage content using PCN.
\end{itemize}

The remainder of this paper is organized as follows. We present the design goals of PCN (\S 2). We provide PCN's basic framework (\S 3) and secure content management methods (\S 4). Then, we present the prototype implementation (\S 5) and the preliminary evaluation results (\S 6). Next, we discuss some of the remaining issues (\S 7). After reviewing the related work (\S 8), we conclude the paper (\S 9).

\begin{figure}
\centering
   \hspace{20pt} \includegraphics[width=0.7\textwidth]{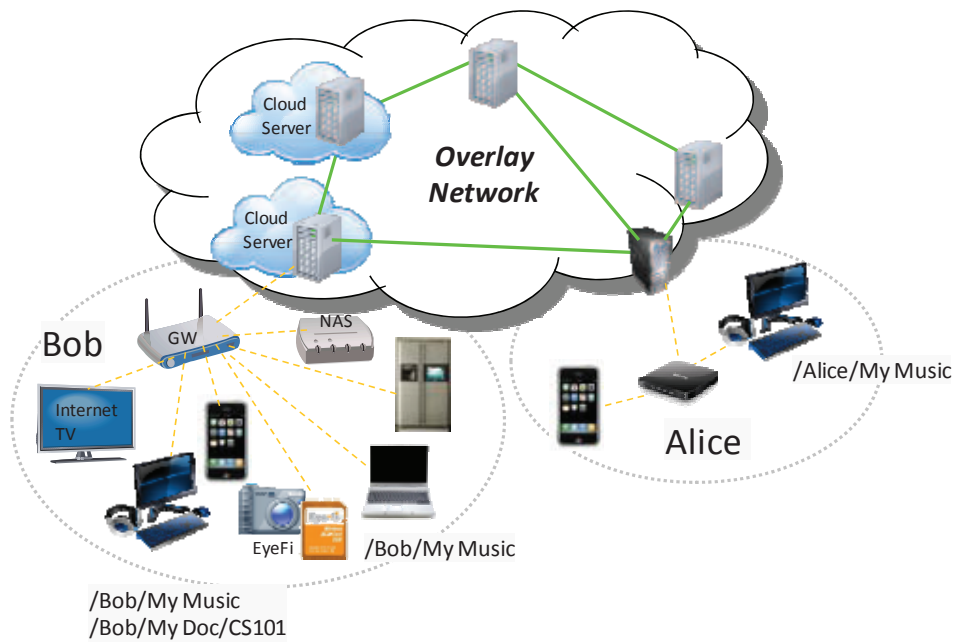}
\caption{Personal content networking scenario: Alice and Bob manage their personal content over multiple devices using single, hierarchical, persistent naming. Bob can share the content with Alice by simply passing the link (say, ``\emph{/Bob/My Doc/CS101}'').} \inv
\label{fig:overview}
\end{figure}

\section{Target Scenario and Design Goals}
We use the following example to motivate the needs of personal content networking. Bob has a number of smart devices (see Figure \ref{fig:overview}): Internet TV, desktop computer, iPhone, WiFi-enabled digital camera, Internet fridge, and network attached storage (NAS). He has other devices at school (e.g. desktop computer, laptop) and also maintains a few cloud servers (e.g. Amazon EC2). His personal content is currently stored across these places, and Bob had a difficult time tracking all these files. For example, his friend Alice asks him to send the lecture material of the course that they took last year. He only remembers that it is located in a document directory, but he forgets which device he put it on. He searches through his devices one by one: laptop, servers, desktop computer, and NAS, and finds that it is stored in his NAS at home. After locating the lecture material, Bob feels a bit frustrated because the size is over 1 GB (even after compression) and he cannot send it via email. He calls Alice saying he will give the files to Alice using a USB stick.

This example clearly illustrates the needs of Personal Content Networking (PCN) such that users can manage personal content stored across multiple devices and selectively share content with intended groups. The design goals of PCN can be summarized as follows:
\begin{itemize}
\item \emph{Single persistent hierarchical namespace}: Single persistent naming of personal content will guide users to have a unified view of their personal content stored across multiple devices. A hierarchical namespace is essential because it has been reported that hierarchical naming significantly lessens the cognitive overhead of locating files~\cite{Lansdale88,Jones05,Dearman08,Henderson09}.
\item \emph{Social networking}: Users often want to share content with their friends. PCN should leverage social networking aspects through establishing and managing trust relationships among friends and through building an overlay network based on these trust relationships for content sharing.
\item \emph{Fine-grained access control}: PCN must provide fine-grained expressive access control to enable secure content management over distributed untrusted servers that store/transfer data properly and yet do not necessarily keep the data confidential.
\item \emph{Disrupted operations}: Because devices can go offline at any time, users should be able to replicate files and files must be automatically synchronized whenever the devices come online again as in existing distributed file systems~\cite{Satyanarayanan90,Reiher94,Mazzola04}.
\item \emph{Security guarantee}: Because PCN manages distributed untrusted devices connected over the Internet, it must be resilient to well-known security attacks such as a denial of service attacks and false data injection attacks.
\end{itemize}
In the above scenario, PCN will allow Bob to easily locate the material (e.g. ``\emph{/Bob/My Doc/CS101}'') and to pass this link to Alice. Bob does not need to examine each device, but simply needs to browse his namespace on any machine. Alice can download the content using the link provided. In the following, we review CCN, which is a building block of PCN (\S 3) and demonstrate how PCN's basic framework can be built over CCN (\S 4).

\section{Basic PCN Framework with CCN}
In this section, we review the core components of CCN: (1) naming, (2) content reachability, (3) content retrieval, and (4) content-centric security, which are the basic building blocks of PCN's underlying content retrieval. We discuss how PCN builds a web of trust and an overlay network based on social relationships, because the key functions of personal content networking is to share content among friends.

\subsection{Background: CCN review}
\emph{Naming}: CCN names a file with a user friendly, structured, effectively location-independent name. Each file is divided into multiple segments. Consider the following example name: \emph{/parc.com/music/abc.mp3/v3/s0}. Here, ``\emph{parc.com}'' is a globally routable name (called a prefix), ``\emph{/music/abc.mp3}'' is a local name in ``\emph{parc.com}'', ``\emph{v3}'' is a version name (represented using a timestamp), and ``\emph{s0}'' denotes the segment number.

\emph{Content reachability}: In CCN, a prefix owner announces their prefix to the entire network. For example, Alice from ``\emph{parc.com}'' announces her files as ``\emph{/parc.com/Alice/}'' from her laptop. Each node in the network broadcasts the incoming prefix to its neighboring nodes. Whenever a node receives the prefix, it establishes a backward pointer to the sender in its Forwarding Information Base (FIB) for that prefix. Content is reachable to the prefix announcer as any content request can be routed by following the backward pointer in the FIB.

\emph{Content retrieval}: Content retrieval is pull-based as in HTTP (i.e. get and response). A user sends a request (via an Interest packet), and any node that has the requested content in its local storage (or cache) can respond to the request. For a given prefix, the Interest packet is forwarded along the reverse path toward each data source through following the backward pointer in the FIB. Whenever a node receives an Interest packet, the breadcrumb information (i.e. the backward pointer to the previous forwarder) is stored in the Pending Interest Table (PIT). Then, the corresponding data packet will be delivered according to the reverse path in the PIT.

When forwarding an Interest packet, CCN uses the longest prefix matching algorithm; i.e. in the FIB, a node finds the longest prefix entry that has the largest number of leading letters matching those of the content name in the Interest packet. For example, Alice's desktop has ``\emph{/Alice/my music/pepper/}'', and Alice's laptop has ``\emph{/Alice/my music/}''. When Bob accesses ``\emph{/Alice/my music/pepper/abc.mp3}'', it matches the prefix entry of ``\emph{/Alice/my music/pepper/}'' and the Interest packet will be delivered to Alice's desktop.

\emph{Content-centric security}: CCN supports secure binding between the name and content (called content-based security) where protection and trust travel with the content itself~\cite{Jacobson09}. To this end, CCN uses asymmetric cryptography: the content is authenticated with digital signatures, and private content is protected with encryption. Each data packet contains the owner's signature $P$, $Sign_P(N,C)$, which covers the name (N) and content (C). This content-based security is critical because content can be cached in untrusted intermediate nodes. For key management, CCN can use a traditional certificate-based public key infrastructure (PKI) or a Web of Trust (e.g. PGP).

\subsection{Naming in PCN}
The current generation of personal devices use rigid and weak naming of the form ``hostname:path''. The key problem is that content is tied to a host, which makes personal content management non-trivial, particularly when a user interacts with a number of devices (e.g. laptop, desktop computer, smartphone) including cloud-based storage services (e.g. Dropbox). A user must track what files are located in which devices/services and decide how to migrate/replicate/update the content.

In PCN, we define a single persistent hierarchical namespace for each person. It is known that global namespaces are politically and technically difficult to implement (e.g. X.509~\cite{x509}, PEM~\cite{pem}). Thus, we use the local decentralized namespaces of SPKI/SDSI~\cite{Clarke01}. Each person has a public-private key pair to verify the identity of the sender (sign/verify) and to ensure privacy (encrypt/decrypt). The relationships among users in PCN are considered to be flat, and it is sufficient to use the public key as an identity. Nonetheless, there are cases where hierarchical naming is useful, e.g. a group of users has a set of sub-groups. In SPKI/SDSI, a user can define a local namespace as follows: a user's key $K$ followed by a single identifier (which is distinct within the local namespace). For example, Alice has a key $K_a$ and makes her own name as ``$K_a$ Alice''. A study group with key $K_g$ can name its sub-groups as ``$K_g$ sub1'' and ``$K_g$ sub2''. If a sub-group has multiple smaller groups inside, that group can name those groups similarly; for example, sub1's two internal groups (ssg1 and ssg2) can be named as ``$K_g$ sub1 ssg1'' and ``$K_g$ sub1 ssg2''. Note that a local name is globally unique because the name contains the public key of the user. Moreover, each user can make signed statements of these \emph{local names}, which allows anyone to certify a key via a web of trust~\cite{Clarke01}.

Given this, the data name has the form $N=P:L$, where $P$ is the user name (or its cryptographic hash), and $L$ is the label representing the location of data in the hierarchy, e.g. Alice's music can be denoted as ``\emph{/$K_a$ Alice/music/}''. PCN's naming can be used in CCN with minimal modification as CCN uses hierarchical naming (e.g. ``\emph{/parc.com/test.txt}''). As in CCN, each device advertises the content reachability information through broadcasting the name prefix of the content that is stored in the device. For example, Bob's laptop will advertise ``\emph{$K_b$ Bob/my doc/}'', and his iPad will advertise ``\emph{/${K_b}$ Bob/my music/Beatles/}''. For the sake of brevity, hereafter we will use \emph{abbreviated names} without a public key, e.g. ``\emph{/Bob/my music/}''.

\subsection{Trust management in PCN}
Each PCN user has a private-public key pair that is used to define their name. When a new device is purchased, this information must be securely installed in order to initialize the PCN service. Moreover, for content sharing with others, a user must establish a trust relationship through securely exchanging public keys (e.g. how does Bob make sure that the key belongs to Alice?). Nonetheless, trust relationships do not necessarily guarantee data confidentiality. For both device initialization and trust establishment problems, secure key distribution is the critical issue. Users can use USB sticks or local/wide area networks for key exchanges. The latter is less secure than the former, because it is vulnerable to man-in-the-middle attacks.\footnote{An attacker can eavesdrop on the channel and make independent connections with the victims and then relay messages between the victims making the victims believe that they are talking directly to each other over a private connection, when in fact the entire conversation is controlled by the attacker.}

A simple method of avoiding the attack is to use another secure channel. Alice can show (or read) her public key to Bob (e.g. via physical presence, SMS, email, voice communications, etc.). She can ask Bob to verify whether his key matches the received key. Given that verifying a large number is laborious and can be erroneous, Ellison et al.~\cite{Dohrmann02} proposed an approach where the keys are represented in color bars so that users can more easily verify the key. In Unmanaged Internet Architecture (UIA), multiple choice questions are used to reduce the user's burden~\cite{Ford06}. For example, Alice sends her multiple choice question to Bob, and Bob sends his question to Alice. After solving each other's question, they exchange the hashed values of their answers (and both keys), hoping that the attacker cannot solve the questions and thus fail to control the conversation.

However, this approach is also vulnerable to man-in-the-middle attacks because a malicious user can perform a dictionary attack. The attacker knows both keys and the multiple choice questions. They can easily find the answer through computing a hashed value for each answer choice and comparing this value with the received answer. Like UIA, we use multiple choice questions, and yet we solve the man-in-the-middle attack using Ellison's approach~\cite{Ellison96}, which is based on Pedersen's interlock protocol~\cite{Pedersen91}. Given that the secret (i.e. the answer of a multiple choice question) is $a$, one chooses a random value $u$ and then computes $x=g^a h^u \mod p$, where $p$ is prime, and $g$ and $h$ are generators of the group $\mod p$. Alice and Bob generate their own numbers and exchange these values, i.e. Alice generates $x_A=g^{a_A} h^{u_A} \mod p$, and Bob generates $x_B = g^{a_B} h^{u_B} \mod p$. The attacker cannot infer the value of $u$ and must use a random value to finish the transaction, which thus effectively thwarts the dictionary attack.

\subsection{CCN overlay construction}
Trust management among friends can be used to form a social network. This social relationship is used to create an overlay network for content-centric networking (CCN). Whenever an identity introduction occurs, the corresponding personal devices also exchange IP addresses and join the overlay network. Each device maintains a peer list that contains the IP addresses and port numbers of other devices. For a given user, the list includes the user's own devices and direct friends' devices. For example, Alice's laptop has a list of all her devices and a list of Bob's devices. These devices periodically check the availability of neighboring devices in order to maintain the overlay network. This is further discussed on a device hidden behind a NAT in (\S 7).

\section{Secure Content Management}
In this section, we first illustrate file replication and synchronization, and we justify the need for prefix protection in replication. Then, we present details about content-centric access control, followed by an illustration of remote content management and a discussion of key revocation.

\subsection{Replication}
PCN supports both file- and directory-level replication services. Replication is straightforward because a user simply needs to republish the fetched content into the local CCN client's repository. Then, the prefix of the file is announced so that the other nodes can also fetch the file. For example, Alice has her favorite song in her laptop and it is located at ``\emph{/Alice/my music/pepper/abc.mp3}''. She simply downloads the file from her desktop computer and asks the local CCN client in her laptop to replicate the file. Now, both her laptop and desktop computers announce the prefix ``\emph{/Alice/my music/pepper/}''. However, directory replication requires more attention because it contains a set of entries each of which associates a name with a pointer to a file or subdirectory. If directory replication is requested, the local client recursively fetches all associated files/subdirectories to the local repository. For example, if Alice replicates ``\emph{/Alice/my music/}'' in her laptop, the local client downloads all files from her remote desktop computer and then announces the prefix ``\emph{/Alice/my music/}''.

Note that in PCN, applications can access files without replicating them. Recall that a CCN node has a two-tier data hierarchy: a local cache (in memory) and a local repository (on a disk)~\cite{Jacobson09}. If a requested file is not present in the local cache, its local repository is examined. If that fails, an Interest packet will be issued and the file will be fetched from a remote node. The fetched file will be temporarily stored in its local cache from which applications can access it. Through doing this, locality can improve the accessibility. Note also that a user should be able to navigate the namespace (e.g. the UNIX \emph{ls} command). By treating a directory as a special file, PCN collects the directory entries from the devices using a procedure similar to that of finding the latest version of a file in CCN.

\subsection{Synchronization}
PCN provides ``eventual consistency'' in that all replicas eventually converge to the same version given sufficient messages exchanged among the participating devices (i.e. a file with the freshest timestamp)~\cite{Reiher94,Mazzola04}. Eventual consistency is a widely used consistency model in disruption-prone mobile environments.

Whenever a replicated file is updated, a new version is created thereafter (timestamp). Each replicated file has an associated version vector that tracks its update history~\cite{Reiher94,Page98}. In order to create an alert for this event, the node that makes the update will re-announce the corresponding prefix with a modification mark, which is a special type of prefix announcement that is used for the update notification. The prefix announcement also contains detailed information of the updated file, including its name, the current version, and the version vector. For synchronization, the local client compares the version vector of the local replica with that of the updated file. If the updated file is strictly newer than the local one, its version vector will dominate; the local client fetches the updated file and replaces the local file in the repository. If two version vectors are not equal and neither one dominates, an update/update conflict occurs. If automatic merging fails, PCN notifies the user that a conflict has been detected. The user will be presented with a revision history including authors, dates, and versioned content. It is left to the user's discretion to resolve the conflicts and mark the content as merged. Note that whenever intermediate nodes hear the modification announcements, their local caches are examined in order to determine whether there are matching files, and the matched files (or data packets) in the caches are invalidated.

Synchronization of a replicated directory needs a special care. Although the modification operations are limited to adding new entries or deleting/changing existing entries, a directory replica can be modified from multiple places, which causes several well-known synchronization issues such as insert/delete ambiguity, remove/update conflicts, and name conflicts~\cite{Balasubramaniam98,Page98}. In PCN, we adopted the existing solutions used in the Ficus file system~\cite{Reiher94,Page98}.

When a node re-joins the PCN network after a disruption, it first checks its neighbors to find any missing prefix announcements. As will be seen, a PCN user has a reserved namespace for devices, namely ``\emph{/dev}'', and devices are accessible through this name, e.g. Alice's iPod is named ``\emph{/Alice/dev/iPod}''. For prefix announcement synchronization, each device stores the received prefix announcements in a designated place, e.g. Alice's iPod has ``\emph{/Alice/dev/iPod/received\_prefix}''. This allows the node to search for the updates of the files located in its local storage. If the node finds a prefix with a modification mark, it performs file synchronization as described earlier.

Note that in PCN, nodes fetch the updated file for synchronization. If the size of a file is large and only a small part of the file is updated, fetching the entire file will waste the bandwidth. A simple solution to this problem is that a node generates/publishes a delta file (e.g. using a diff file) and includes a link to the delta file in the prefix announcement.

\subsection{Prefix protection}
Thus far, we have assumed that any node can replicate the content and announce the name prefix. After replicating the content, however, malicious users can launch an attack through inundating the network with fake update announcements. PCN nodes could waste considerable resources managing these fake updates. Given that CCN does not manage updates, this problem is unique to PCN. In order to manage this, we impose a restriction that the prefix announcement must be signed by the prefix owner. This technique is a reasonable approach because people typically want to have full control of their namespace and the locations of files in a multi-device environment. A similar technique is used in a secure BGP where each prefix is signed in order to prevent prefix hijacking where an attacker has partial or full control of the named prefix~\cite{Kent00}.

In PCN, a prefix announcement is augmented to include a signature that certifies the prefix ownership. Furthermore, we implement the ownership delegation such that an owner certifies that a named user is allowed to announce the named prefix through issuing a prefix certificate. For example, Alice can issue a certificate to Bob that he can announce the prefix ``\emph{/Alice/my doc/proj/}''. Intermediate nodes can verify that the certificate is valid and that the prefix announcement originates from Bob (similar to data packet validation). Bob can update his local replica of Alice's file, and the update will be automatically propagated. Note that it is possible for the attackers to perform a replay attack where a CCN speaker replays a prefix that it has previously heard. This problem can be mitigated through adding an expiration timer as in S-BGP~\cite{Kent00}.

\subsection{Content centric access control}
Access control in personal devices is primarily host-centric. In Identity-Based Access Control (IBAC)~\cite{Sandhu94}, a user first logs into the system (authentication) and then accesses files based on the permissions in the access matrix (authorization). SPKI/SDSI supports Role-Based Access Control (RBAC) where permissions in the access matrix are tied to roles~\cite{Clarke01}. SPKI/SDSI is also host-centric because it assumes trusted servers and insecure channels, i.e. an individual must first set up a secure channel (using SSL) to prevent man-in-the-middle attacks, and the server verifies whether a requester's key is on the role-based ACL~\cite{Burnside02}. However, in PCN, host-centric access control is not suitable because the channel is not secured but the data itself is secured.

In PCN, we need \emph{content-centric access control}, i.e. the access control of content is self-contained and is not tied to a host. A simple solution is to encrypt the content using the receiver's public key and to define a specific name for the encrypted data that is meaningful to the receiver. The encrypted content can be placed on untrusted servers because others cannot decrypt the content. If a file needs to be shared with multiple people, a common key is used for the file encryption, and this key is encrypted using each user's public key~\cite{Goh03}. Then, the encrypted keys are included in the metadata of an encrypted file, and the entire content (metadata + encrypted file) is published. However, this approach has several limitations. Supporting expressive access control is difficult because it only uses a single file encryption key. If multiple keys are used, the metadata size linearly increases with the number of users/groups. More importantly, once the content is published, the owner cannot give access to other users; that is, the owner must republish the original file and include additional users.

In PCN, we solve this problem using Ciphertext-Policy Attribute Based Encryption (CP-ABE), which permits secure sharing of content within a group across multiple untrusted servers~\cite{Bethencourt07}. ABE is the key enabler for Attribute-Based Access Control (ABAC) where access decisions are based on the attributes associated with individuals. First, each user generates an ABE public key (PK) and an ABE master key (MK). A user can define a set of attributes (e.g. college friends, CS219 team, family members, etc.) and an access policy using Boolean formula for the attributes. This allows a user to perform fine-grained, expressive access control. The user assigns a set of attributes to each user and then issues a secret key corresponding to the attribute set, i.e. \emph{Secret Key (SK) Generation(MK, S)}, where $MK$ is the master key and $S$ is a set of attributes assigned to the user. A file can be encrypted using the public key and access policy, i.e. \emph{Encrypt(PK, M, A)}, where $PK$ is the public key, $M$ is a message, and $A$ is an access policy. Here, any user can encrypt the file using the public key. Furthermore, any user who has a secret key with attributes that satisfy the policy can decrypt the content, i.e. \emph{Decrypt(PK,CT, SK)}, where $PK$ is the public key, $CT$ is the ciphertext, and $SK$ is the secret key~\cite{Bethencourt07}. In ABE, the metadata size does not linearly increase with the number of users/groups because its metadata only contains the access policy information whose size scales with the number of attributes. Moreover, the owner can still issue attribute keys for published content without republishing the content.

Assume that Alice would like to selectively share her music collection, ``\emph{/Alice/my music/rock}.'' This content is digitally signed using Alice's publisher key and is published under that namespace. For access control, ``\emph{Red Hot Chili Peppers}'' is encrypted with the attribute ``\emph{college friends}.'' ``\emph{Incubus}'' is encrypted with both ``\emph{college friends}'' and ``\emph{CS219 team}'' attributes because Alice discussed Incubus with her teammates and wants to share the songs with them only. Bob is Alice's college friend, and Alice issues a secret key for the attribute ``\emph{college friends}''; Cathy is Alice's CS219 team mate, and Alice issues secret keys for the attributes of ``\emph{college friends}'' and ``\emph{CS219 team}.'' Bob can decrypt ``\emph{Red Hot Chili Peppers},'' while Cathy can decrypt both ``\emph{Red Hot Chili Peppers}'' and ``\emph{Incubus}.''

\begin{figure}
\centering
  \includegraphics[width=0.6\textwidth]{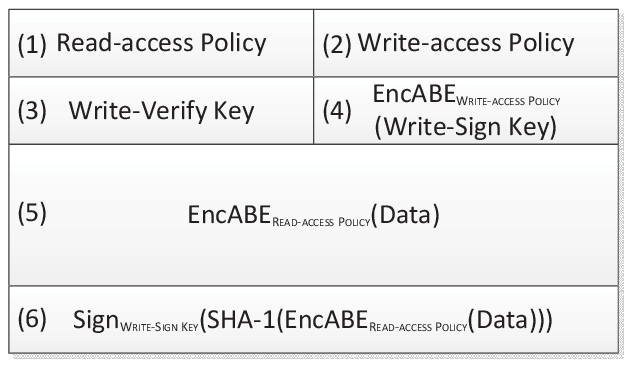}
\caption{PCN's file data structure for secure content centric access control}
\label{fig:ccac}
\end{figure}

In PCN, an owner of a file can set access permissions of \emph{read} and \emph{write} using separate access policies (as used in ABE). The resulting access modes in PCN are \emph{read-only} and \emph{read-write}; the write-only mode is not suitable for personal content networking. Access modes can be also used with directory files in order to limit access of a directory listing. As shown in Figure \ref{fig:ccac}, the PCN payload contains the following fields: (1) read-access policy, (2) write-access policy, (3) write-verify key (public key), (4) write-sign key encrypted using ABE with write-access policy, (5) actual data encrypted using ABE with read-access policy, and (6) write signature (optional). For write access control, a file owner issues a private-public key pair that is located in fields (3) and (4). The write-sign key is only accessible to those who have write permission because the write-sign key is encrypted using ABE with the write-access policy. Whenever a file is updated, the file is encrypted using ABE with the read-access policy. This legitimate modifier then reads the write-sign key through which it generates the signature of the updated content, which is placed in field (6). Then, this update event is notified to all nodes that replicate the content via a prefix announcement with a modification mark. The replica nodes will then fetch the updated content and verify whether it is modified by legitimate users who satisfy the write-access policy. Note that in our prototype implementation, we use symmetric encryption in order to reduce the overhead of encrypting/decrypting the content, i.e. the content is encrypted using AES, and ABE is used to encrypt the AES key.

\subsection{Replica management}
A user may want to know what files are stored where and wish to replicate files to remote devices. Regular content browsing such as the UNIX command $ls$ does not tell users in which device the files are located. For replica management, PCN reserves a special directory for devices, namely the ``\emph{/dev}'' directory through which a user can freely name their personal devices. For example, Alice's iPad can be named ``\emph{/Alice/dev/iPad}''. Furthermore, each device announces its device name prefix in device-to-device communications over the CCN.

Similar to the device communications through files in UNIX, the user can write a replication command to a reserved device file, e.g. ``\emph{/Alice/dev/iPad/.cmd}''. After updating the file, its prefix will be announced to the network, and the target device is notified of the update. Then, the device will synchronize the file by fetching the most up-to-date ``\emph{.cmd}'' file. The device finally executes the replication command as specified in the file. Note that for security purposes, PCN restricts this function so that only the prefix owner can update the device files, and the files should be signed using the owner's private key. Multiple concurrent requests can be managed using serialized updates based on timestamps. Delayed execution is not permitted and requests expire after a threshold period of time.

\subsection{Key revocation}
PCN primarily uses the following keys: a personal public-private key pair, group public-private key pairs, and ABE keys. Secure key distribution can be assured because PCN uses the secure identity exchange mechanism and relies on an SPKI/SDSI-style web of trust. Any intermediate nodes will be able to correctly acquire public keys that are then used to validate the secure binding between the name and data. While we can leverage CCN for secure key distribution, we must be able to appropriately handle key revocation scenarios for a public-private key pair and an ABE attribute secret key.

If a user's public-private key pair is compromised, the existing key pair can be revoked through the prefix announcement with a revocation mark that is similar to a suicide note in PGP~\cite{Rivest98}. Recall that in CCN, we added two additional prefix types of modification (for update notifications) and revocation (for revocation notifications). Then, the user will generate a new key pair and distribute the public key via the secure identity introduction process, which guarantees that attackers cannot impersonate potential victims. Note that the same procedure can be used to manage the case where a group's public-private key pair is compromised.

If an ABE attribute secret key is compromised or the owner wants to revoke a specific attribute, the owner must revoke both the master key and public key because CP-ABE does not provide a mechanism for revoking an individual attribute. While CP-ABE has a single attribute revocation through the addition of a timer attribute for each attribute, this approach is less practical because it complicates the overall system: (1) the owner must periodically issue keys and all files must be re-encrypted with new attribute sets, and (2) a tamper-proof clock is required to ensure the security guarantee.

Whenever the master key and public key are revoked, the owner must re-encrypt all files. However, this process is very expensive. In order to reduce the overhead, we employ the lazy revocation scheme proposed by Kallahalla et al.~\cite{Kallahalla03}. Unlike the compromise of a public-private key pair, that of an ABE attribute secret key is less serious, as long as the revoked user (or attacker) only has read-only access rights: the revoked user cannot remove or update files. In this scenario, it can generally be assumed that the revoked user has read and copied all files, and it is still acceptable for the user to read unmodified or cached files. However, the lazy revocation ensures that the revoked users are not able to read \emph{updated files}; that is, the updated files will be re-encrypted with the new ABE public key.

Note that it is also possible for users to immediately revoke all keys and re-encrypt all files. In this case, the user must undergo a series of steps: (1) re-generate a new ABE key set, (2) invalidate all cached files via prefix announcement, (3) remove the replicas from multiple devices, and (4) re-encrypt all files and re-distribute the replicas.

\section{Related Work}
\emph{Distributed Peer-to-Peer file systems}: Research on distributed file systems for mobile environments has been primarily directed toward extending the existing client/server-based file systems to manage node mobility and network disruption~\cite{Reiher94,Sobti02,Mazzola04}. A common technique is to use optimistic file replication and eventual consistency. BlueFS~\cite{Nightingale04} extends the client/server-based file system through focusing on power management to save energy in mobile devices. Compared with coda (i.e. a distributed file system), BlueFS substantially reduces the file system energy usage and provides up to three times faster access to data replicated on portable storage. EnsemBlue~\cite{Peek06} builds upon BlueFS and provides a consistent view of all files located across multiple devices with heterogeneous device capabilities in a self-organized manner. EnsemBlue supports namespace diversity through translating between its distributed namespace and the local namespaces of consumer electronic devices. It further supports extensibility through persistent queries, which is a robust event notification mechanism that leverages the underlying cache consistency protocols of the file system. Ficus~\cite{Reiher94} uses a flexible peer-to-peer (P2P) model for optimistic replication where all replicas are equal and can propagate updates to all other replicas. It has also been reported that Ficus reliably detects all possible conflicts. Bayou~\cite{Terry95} is a P2P weakly consistent storage system where clients are able to connect to any available server to perform reads and writes. In order to support automatic conflict detection and resolutions, it uses anti-entropy for consistent management and supports a database language for data retrieval. Ivy extend a multi-user read/write P2P file system~\cite{Muthitacharoen02}. A detailed survey of P2P file sharing has been presented in this survey paper~\cite{Chothia05}.

Several systems have been designed for multi-device environments. Unmanaged Internet Architecture (UIA) provides zero-configuration connectivity among mobile devices through personal names~\cite{Ford06}. Unlike the existing work, UIA assumes that each device has its own persistent namespace, and a user must track all files located across multiple devices. In contrast, Eyo~\cite{Strauss09} offers a device transparency model in which users view and manage their entire data collection of all devices through periodically flooding metadata everywhere. PersonalRAID~\cite{Sobti02} supports optimistic replication at a volume level, and a mobile storage device is used to provide the abstraction of a single coherent storage name space that is available everywhere, and it ensures reliability through maintaining data redundancy on a number of storage devices. Footloose~\cite{Mazzola04} is a user-centered data store that can share data and reconcile conflicts across diverse devices. Footloose supports application-specific optimistic replication with eventual consistency (e.g. address books), and yet it uses a persistent flat namespace (called ObjectID).

\emph{Wide area P2P storage systems}: Wide area P2P storage can be classified based on the overlay structure: (1) a structured system (e.g. PAST~\cite{Druschel01}, CFS, Ivy) forms a structured overlay network using a distributed hash table (DHT), and (2) a structure-less scheme (e.g. Gnutella and eDonkey) forms a structure-less overlay network where the overlay links are arbitrarily established. Unlike unstructured P2P networks, DHTs provide better performance for searching for items over a large number of distributed nodes, and they have been widely adopted to implement wide area P2P storage. Most P2P storage systems assume wired Internet scenarios and support strong consistency, which is less suitable for personal content networking. As a recent work, Plethora focuses on semi-static peers with strong network connectivity and a partially persistent network state. In a semi-static P2P network, peers are likely to remain participants in the network over long periods of time (e.g., compute servers), and are capable of providing reasonably high availability and response-time guarantees~\cite{Ferreira05}.

\emph{Decentralized access control}: The following concepts are closely related: user authentication, access control authorization, and data confidentiality. Existing access control systems can be classified based on their authentication method. When AUTH\_SYS (UNIX's default) and Kerberos are used, systems mostly provide UNIX-style ACL (e.g. Network File System (NFS), Andrew File System (AFS), xFS~\cite{xfs}). When public key cryptography is used, systems typically support either UNIX-style ACL (e.g. SFS~\cite{Mazieres99}) or certificate authorization (e.g. DisCFS~\cite{DisCFS}). These systems assume that \emph{file servers are trustworthy}, but the network is not secure; thus, data confidentiality is guaranteed through securing the channel (e.g. SSL). If the servers are not trustworthy, we can either rely on other semi-trusted servers as in Cobalt~\cite{Veeraraghavan08} or use cryptographic encryption to preserve data confidentiality as in Cryptographic File System (CFS)~\cite{Blaze93}, Plutus~\cite{Kallahalla03}, and SiRiUS~\cite{Goh03}. A detailed survey of recent decentralized access control has been presented in this survey paper~\cite{Miltchev08}.

In the CFS approach, authentication is typically undertaken using public key cryptography where a user's public key is used as an ID, and a digital certificate is used for authentication. CFS uses a single key for encryption (coarse-grained, e.g. directory/volume) and is dependent on the underlying file system for write authorization~\cite{Blaze93}. Later variants use a lockbox to protect the keys (with more fine-grained access control) and introduce several mechanisms for verifying the write operations without depending on the underlying file system~\cite{Goh03,Kallahalla03}. In particular, SiRiUS~\cite{Goh03} assumes that the network storage is untrusted and provides its own read-write cryptographic access control for file-level sharing. It permits a file to be shared by multiple individuals or groups using a common file encryption key that is encrypted again using each user/group's public key.

Given that Attribute-Based Encryption (ABE) is designed to provide fine-grained, expressive access control, several existing works have used ABE for \emph{read-only} content sharing over untrusted storage~\cite{Yu08,Baden09,Yu10}. In particular, Yu et al.~\cite{Yu10} used a key policy ABE (KP-ABE) to provide privacy-aware content sharing over untrusted cloud storage and Proxy Re-Encryption (PRE) to delegate the task of re-encryption on cloud servers. While PCN is considered to be a cryptographic file system, unlike existing systems, PCN provides fine-grained expressive access control using CP-ABE in a fully distributed environment with untrusted nodes and it allows file owners to set up expressive \emph{read and write} access policies based on attributes (e.g. college friends, family members, etc.). Furthermore, none of the aforementioned systems provide \emph{secure binding} between the name and data; thus, the channel must be secured in order to prevent man-in-the-middle attacks.

\section{Prototype Implementation}
\begin{figure}
\centering
  \includegraphics[width=0.6\textwidth]{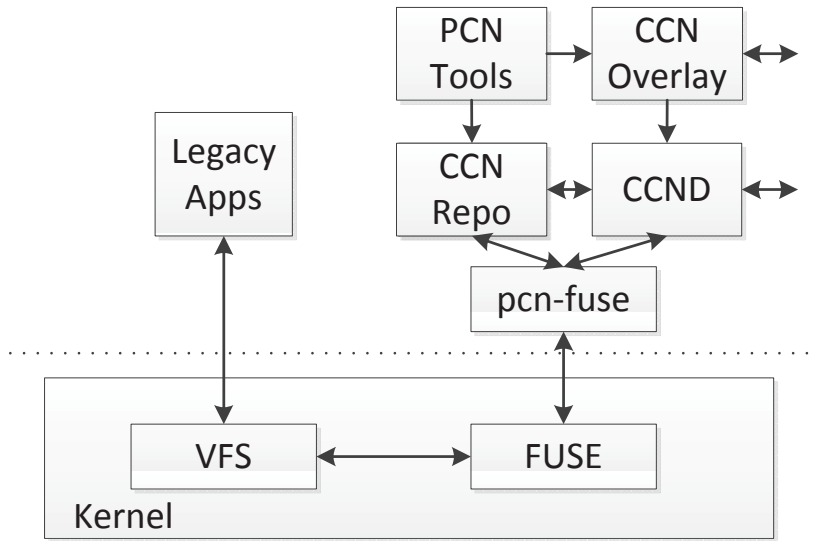}
\caption{PCN architecture} \inv\inv
 \label{fig:arch}
\end{figure}

We implemented a PCN prototype in the Linux and Android platforms, and further integrated the prototype with FUSE, which is a user-level file system to support legacy applications in Linux. As shown in Figure \ref{fig:arch}, the PCN tools include \emph{pcn-init}, \emph{pcn-intro}, \emph{pcn-abe-enc}, and \emph{pcn-broswer}. The \emph{ccn-overlay} tool maintains the CCN overlay network. The \emph{pcn-fuse} tool implements the basic VFS file operations, which support legacy applications.

\begin{figure}
\centering
  \includegraphics[width=0.5\textwidth]{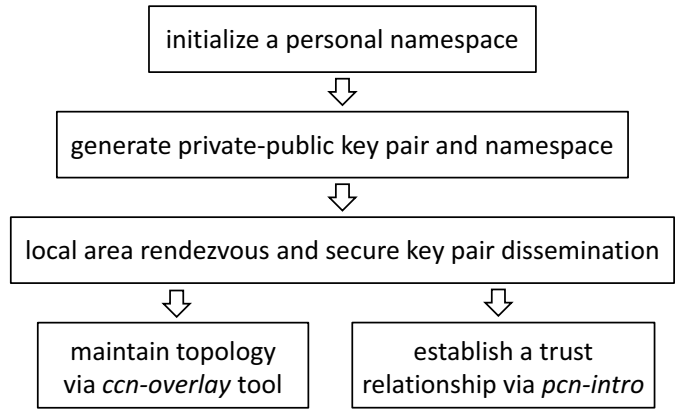}
\caption{Flowchart of topology and trust relationship establishment} \inv\inv
 \label{fig:arch}
\end{figure}

As depicted in Figure \ref{fig:arch}, a PCN user first initializes a personal namespace using the \emph{pcn-init} tool. This tool generates a private-public key pair and prompts the user to name their namespace. The key pair must be securely disseminated to the other personal devices. The tool runs a local area rendezvous tool to locate other devices on the local area network and installs the key pair securely. The device information will be reported to the CCN overlay client that configures its local CCN daemon (CCND)~\cite{ccnx}. Because the current CCNx codebase only supports manually configured, static network topologies, we implemented an overlay network client (called \emph{ccn-overlay}) that builds and maintains an overlay network based on social relationships. Whenever the network topology changes, an overlay client uses external commands to reconfigure the local CCND. The prefix announcement is disseminated through the overlay clients because the current CCNx codebase does not fully support the prefix announcement feature. Each client periodically exchanges ping messages to verify whether its neighboring nodes are alive. Recall that we have three types of prefix announcements, i.e. regular prefix, modification, and key revocation announcements. Besides the device initialization, users can establish a trust relationship using the \emph{pcn-intro} tool, which is based on UIA's device management UI tool~\cite{Ford06}.

In PCN, non-privileged users can mount a namespace into their local user directories, and the legacy applications can seamlessly access the files in the PCN namespace. To this end, we use FUSE, which is a loadable kernel module that allows non-privileged users to create their own file systems without editing the kernel code by running file system operations in the user space while the FUSE module provides a bridge to the actual kernel interfaces. We implemented key virtual file system (VFS) operations in the \emph{pcn-fuse} tool: \emph{getattr}, \emph{getdir}, \emph{mkdir}, \emph{rename}, \emph{open}, \emph{release}, \emph{read}, and \emph{write}. In particular, when an application reads a file, our user-level module downloads chunks through CCN. When a file is modified, we restrict the modification to being committed to its local repository (a new version is created) only if the file is finally released. Then, the \emph{pcn-fuse} tool sends the prefix announcement with a modification mark via the \emph{ccn-overlay} daemon.

\begin{figure}
\centering
  \includegraphics[width=0.7\textwidth]{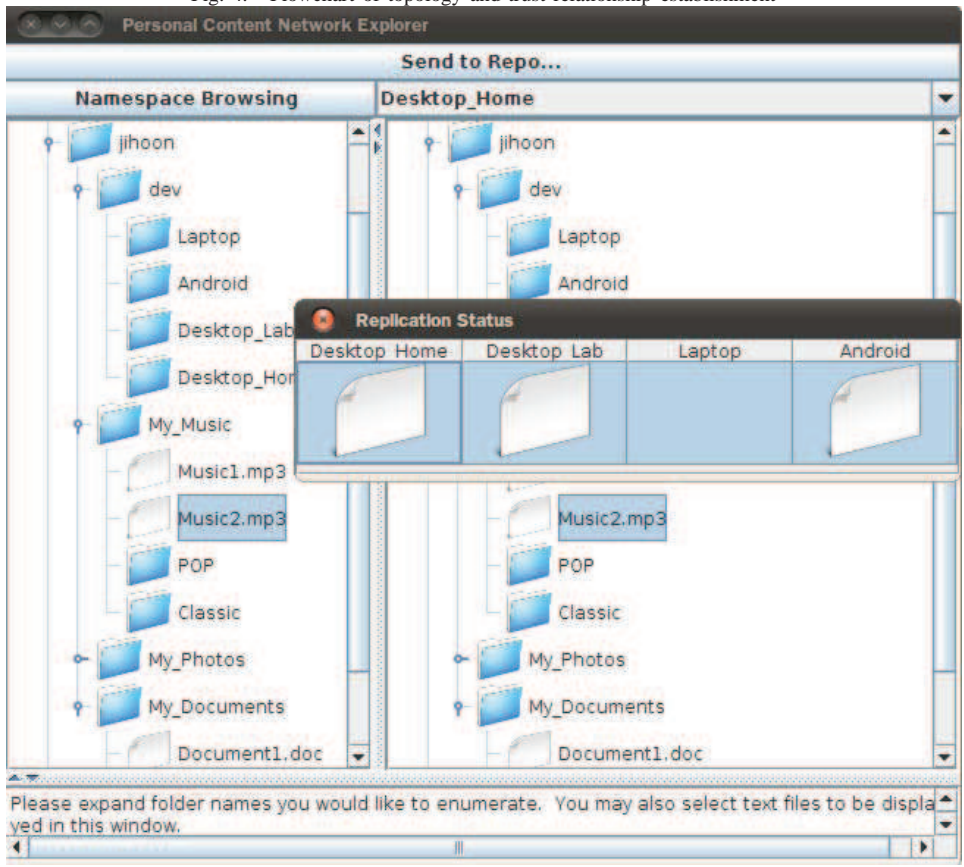}
\caption{Content management tool (\emph{pcn-browser}). The left panel is a namespace browser, and the right panel is the replicated namespace at a specific device (\emph{Desktop\_Home}). By right clicking the file (\emph{Music2.mp3}), a user can see the replication status across different devices.}
 \label{fig:browser}
\end{figure}

For content management, we implemented \emph{pcn-broswer} through significantly modifying the \emph{ccnbrowser} tool in the original CCNx codebase (see Figure \ref{fig:browser}). The \emph{pcn-browser} tool allows users to browse any namespace and also displays the file locations. A user can easily issue a replica management command through selecting a file/directory and a target device.

For ABE support, we used the CP-ABE toolkit~\cite{cpabe}. A file published in the local repository can be encrypted using \emph{pcn-abe-enc}. This tool communicates with the local CCN repo daemon, encrypts the named file, and re-publishes the file into the repository. The ABE keys are stored in a user's local keystore (e.g. \emph{.ccnx} at home). If a file is encrypted, the \emph{ccn-fuse} tool automatically decrypts the file and returns the plaintext to the reader. It accesses the user's local keystore for decryption. We ported the CP-ABE toolkit to the Android platform via cross-compilation. Because the CCNx codebase supports the Android platform, we integrated the basic PCN tools into the mobile platform.

\begin{table}[t]\footnotesize
\begin{center}
 \begin{tabular}{ | c | c | c | c | c|}
        	\hline
        	   &  MK setup  &  SK: 5   &  SK: 10 &   SK: 15  \\
        	\hline
        	\hline Laptop  &  166($\pm$0.2)  &  531($\pm$0.4)  &  913($\pm$0.2)  &   1343($\pm$1.9)   \\
        	\hline Mobile  &  354($\pm$0.9)  &  2068($\pm$0.5)  &  3981($\pm$0.5)  &   5947($\pm$0.3)   \\
        	\hline \end{tabular}
\caption{CP-ABE performance of Laptop (L) and Nexus One (M) in milliseconds: master key (MK) setup and secret key (SK) generation with $k$ number of attributes}\inv\inv
\label{tb:cpabe}
\end{center}
\end{table}

\begin{table*}[t]\footnotesize
\begin{center}
 \begin{tabular}{ |   c | c | c | c | c | c | c | }
               \hline
                 & 1KB & 10KB & 100KB & 1MB & 10MB & 100MB \\
               \hline  [D2L] CCNx retrieve & 1152 ($\pm$0.4) & 1225.8($\pm$1.0) & 1410($\pm$2.5) & 2102.2($\pm$2.3) & 11085.4($\pm$16.3) & 80593.8($\pm$139.1) \\
               \hline  Local ABE pri-key & 8.2($\pm$0.1) & 10.2($\pm$0.1) & 9.8($\pm$0.) & 9.4($\pm$0.2) & 13.6($\pm$0.1) & 16.6($\pm$0.1) \\
               \hline  Remote ABE pub-key & 358($\pm$0.1) & 343.6($\pm$0.1) & 346($\pm$0.2) & 348.6($\pm$0.2) & 346.4($\pm$0.2) & 348.4($\pm$0.1) \\
%               \hline  AES-256 content encrypt & 2.1($\pm$0.1) & 2.6($\pm$0.1) & 3.0($\pm$0.1) & 36.1($\pm$0.3) & 331.7($\pm$0.4) & 3925.2($\pm$2.1) \\
               \hline  AES key decrypt & 37.8($\pm$0.3) & 37.7($\pm$0.2) & 37.8($\pm$0.4) & 36.8($\pm$0.9) & 37.3($\pm$0.5) & 37.1($\pm$0.5) \\
               \hline  Content decrypt & 1.0($\pm$0.1) & 1.0($\pm$0.1) & 3.2($\pm$0.1) & 35.0($\pm$0.2) & 380.2($\pm$1.6) & 3946.7($\pm$0.8) \\

               \hline\hline [L2M] CCNx retrieve & 784.8($\pm$0.5) & 973.6($\pm$0.9) & 1157.8($\pm$0.5) & 2273.2($\pm$4.3)  & 10751($\pm$18.5) & 106752.6($\pm$154.6) \\
               \hline Local ABE pri-key & 37.6($\pm$0.1) & 39.6($\pm$0.1) & 38.6($\pm$0.0) & 37.6($\pm$0.1) & 39($\pm$0.0) & 39.4($\pm$0.1) \\
               \hline Remote ABE pub-key & 527($\pm$0.5) & 533($\pm$0.2)  & 532.2($\pm$0.2) & 536.4($\pm$0.2) & 538.8($\pm$0.1) & 539($\pm$0.1) \\
%               \hline  AES-256 content encrypt & 3.0($\pm$0.1) & 18.1($\pm$0.1) & 95.3($\pm$0.2) & 491.5($\pm$1.0) & 3246.4($\pm$3.8) & 21799.2($\pm$5.0) \\
               \hline  AES key decrypt & 425.8($\pm$0.4) & 428.6($\pm$0.3) & 428.1($\pm$0.8)& 427.1($\pm$1.3) & 425.6($\pm$3.3) & 433.6($\pm$6.1) \\
               \hline  Content decrypt & 2.1($\pm$0.1) & 19.9($\pm$0.3) & 120.0($\pm$0.2) & 402.0($\pm$1.1) & 3414.9($\pm$2.3) & 20713.4($\pm$5.3) \\

               \hline\hline [D2M] CCNx retrieve & 706.4($\pm$0.1) & 914.2($\pm$0.6) & 1146.2($\pm$0.3) & 2096.2($\pm$1.1)  & 10259.4($\pm$26.6) & 96724.2($\pm$218.5) \\
               \hline Local ABE pri-key & 35($\pm$0.1)  & 36.2($\pm$0.1)  & 37.6($\pm$0.1)  & 38.6($\pm$0.1)  & 39.2($\pm$0.1)  & 39.6($\pm$0.1)  \\
               \hline Remote ABE pub-key & 532.4($\pm$0.1) & 425.8($\pm$0.1)  & 433($\pm$0.1) & 432.4($\pm$0.1) & 431.8($\pm$0.1) & 431.4($\pm$0.1) \\
%               \hline  Symmetric content encrypt & 3.2($\pm$0.1) & 17.7($\pm$0.1) & 94.9($\pm$0.2) & 486.1($\pm$1.0) & 3209.4($\pm$3.5) & 21581.2($\pm$5.0) \\
               \hline  AES key decrypt & 427.1($\pm$5.3) & 419.6($\pm$7.1) & 429.7($\pm$7.3)& 435.1($\pm$10.1) & 429.3($\pm$6.1) & 435.8($\pm$11.3) \\
               \hline   Content decrypt & 2.4($\pm$0.1) & 18.1($\pm$0.4) & 128.0($\pm$0.1) & 387.1($\pm$1.3) & 3371.1($\pm$2.5) & 21001.4($\pm$4.1) \\
                       \hline
\end{tabular}

\caption{Breakdown of retrieval time (in milliseconds) of a file. D2L: Desktop computer to Laptop; L2M: Laptop to Nexus One; D2M: Desktop computer to Nexus One. Each result is the mean of five trials with a 95\% confidence interval. Each trial was run by setting the CCN cache size to 0 (CCND\_CAP=0) and restarting the CCND between each run in order to reset the local cache.}

\label{tb:d2dm}
\end{center}
\end{table*}

\section{Evaluation}
We present our preliminary system evaluation that answers the following questions: (1) What is the overhead of ABE? (2) What is the detailed performance of each component used in PCN? (3) Given realistic user traces, what is the overhead of PCN (e.g. routing table size, update overhead, etc.)?

In order to provide secure personal sharing, we designed our implementation to incur minimal overhead to the existing CCNx codebase. While a complete evaluation of the CCNx method traces is outside the scope of this paper, our experience demonstrates that the CCNx performance improves with every release. We analyzed the performance of providing security using ABE in the three major areas of (1) key setup and generation, (2) encrypting and storing content, and (3) retrieving and decrypting content. In order to model user behavior, we measured the performance using a mobile device (the Android Nexus One from Qualcomm Snapdragon with 1 GHz CPU and 512 MB of RAM), a laptop (Dell Inspirion 9400 with Intel dual core 2 GHz CPU, 2 GB of RAM, and Intel WiFi Link 5300 that runs Ubuntu 10.10 with Linux 2.6.35), and a desktop computer (Apple iMac with Intel i5-2500s, 2.7 GHz CPU, and Broadcom Gigabit Ethernet that runs Ubuntu 10.10 with Linux 2.6.35). The measured device-to-device TCP performance using Iperf is given as follows (over an average of 10 trials with a 95\% confidence interval): Nexus One to laptop over WiFi: 8.21 Mbps ($\pm$0.02), laptop to Nexus One over WiFi: 8.00 Mbps ($\pm$0.01), desktop computer (wired) to laptop (WiFi): 10.34 Mbps ($\pm$0.02).

\begin{figure*}[t]
\begin{center}
  \begin{minipage}[t]{.49\textwidth}
	\centering
    \includegraphics[angle=0,width=\textwidth]{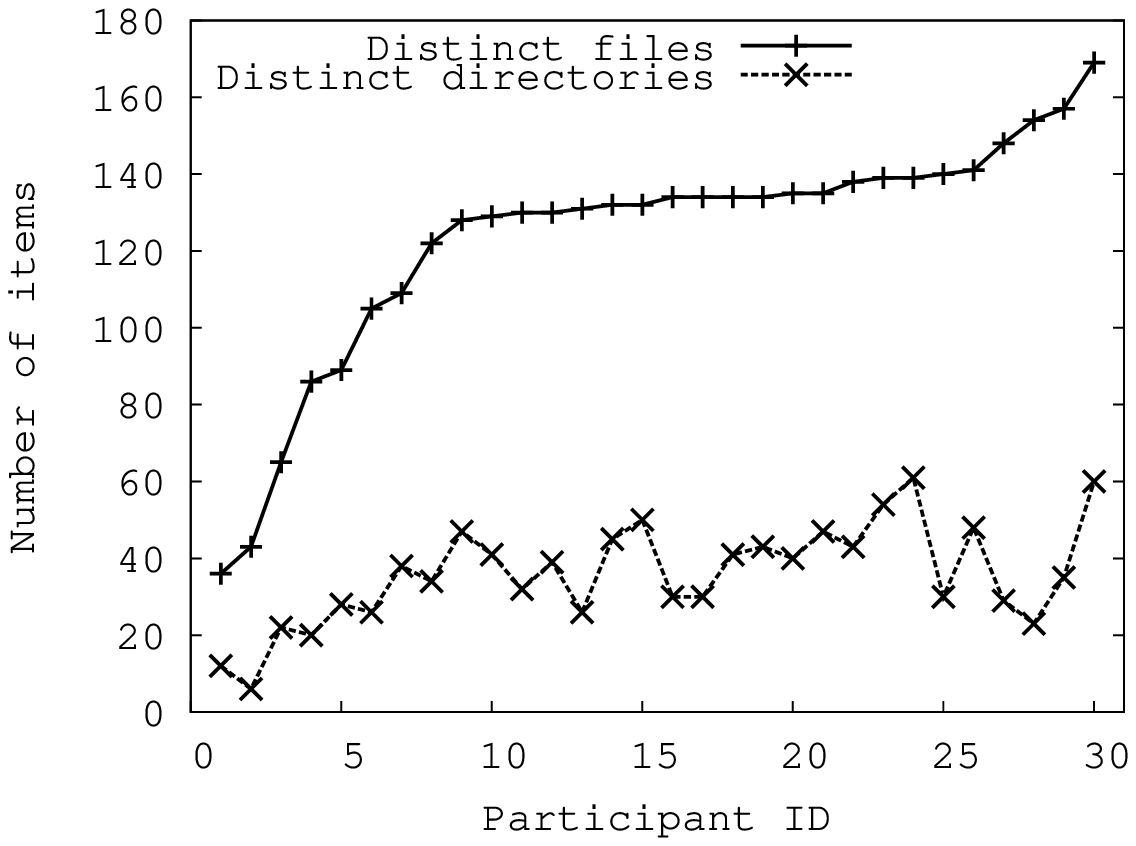}\inv
	\caption{Recently accessed files and directories}
	\label{fig:access}
\end{minipage}
  \begin{minipage}[t]{.49\textwidth}
	\centering
    \includegraphics[angle=0,width=\textwidth]{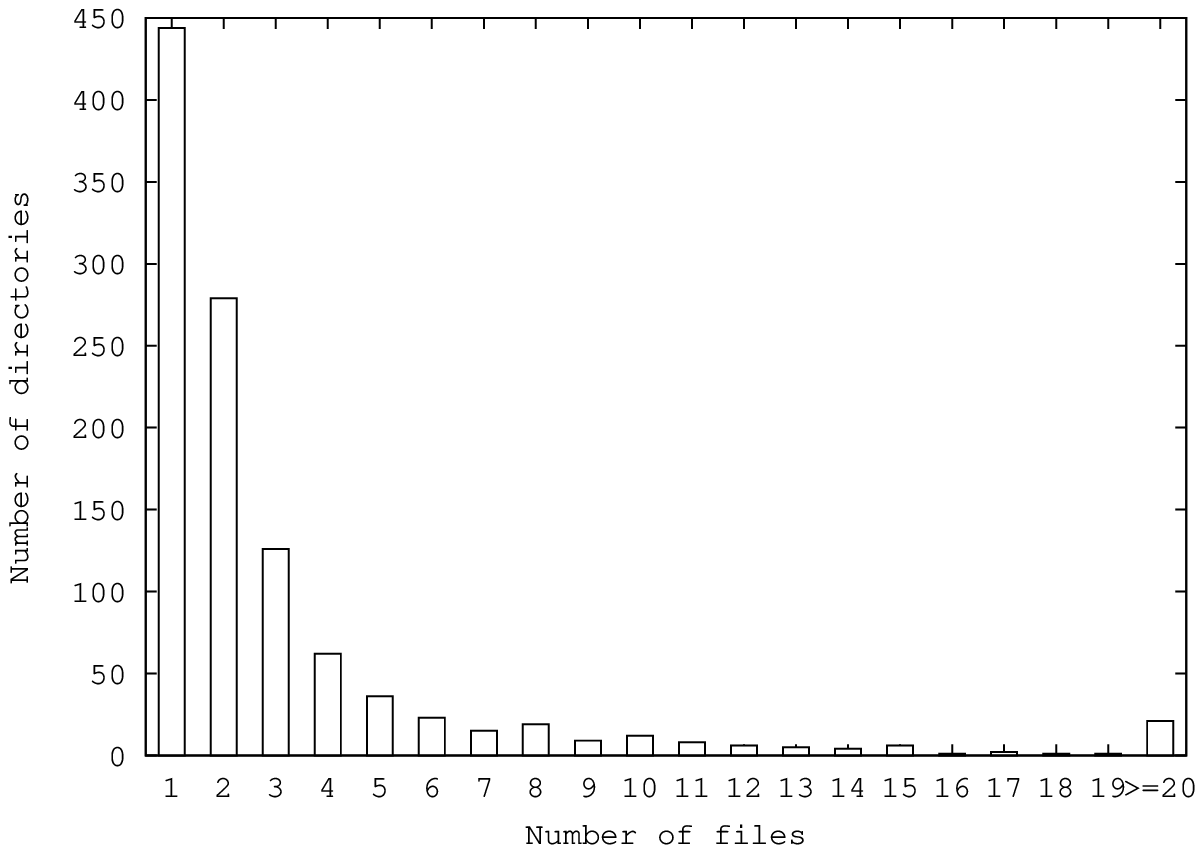}\inv
	\caption{Number of files accessed per directory}
	\label{fig:dir}
  \end{minipage}
\begin{minipage}[t]{.49\textwidth}
	\centering
	\includegraphics[angle=0,width=\textwidth]{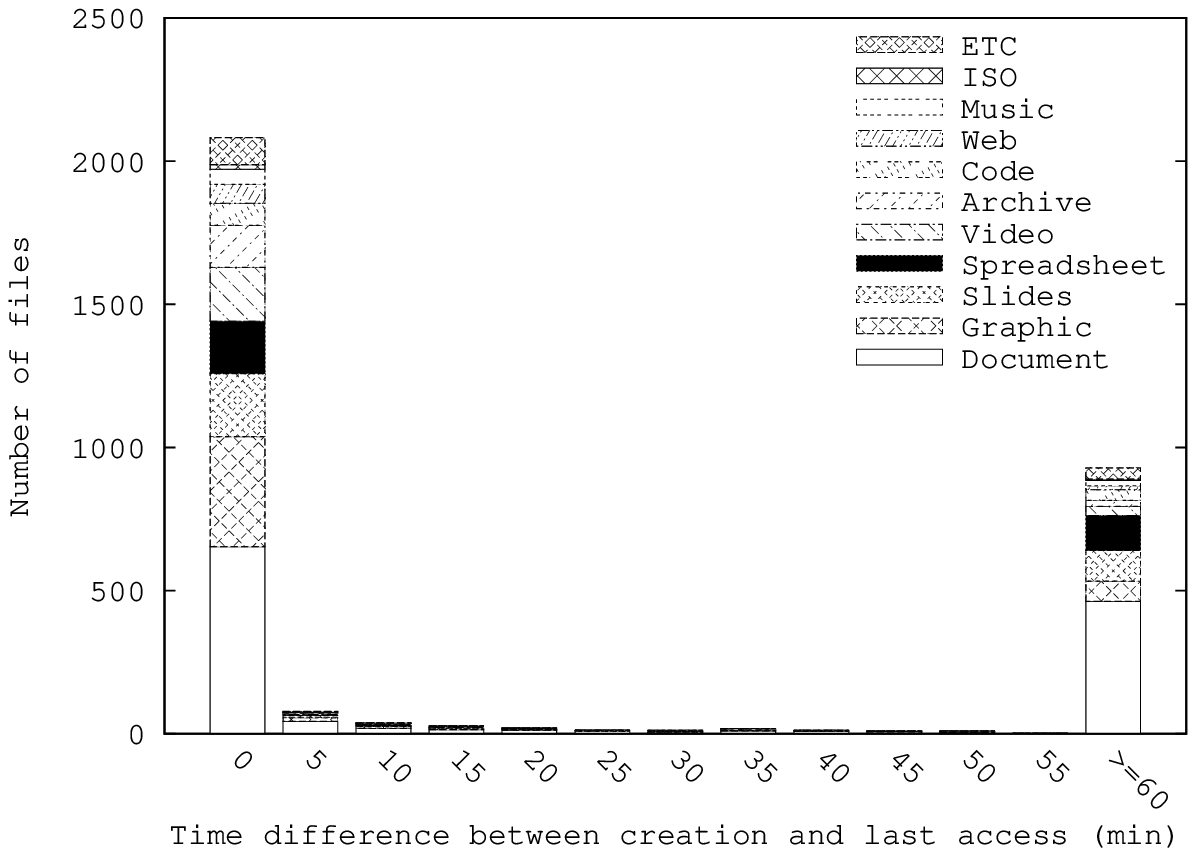}\inv
	\caption{Distribution of access time span based on file types}
	\label{fig:delay}
  \end{minipage}
\end{center}
\inv\inv\inv
\end{figure*}

\emph{Overhead of ABE}: Table \ref{tb:cpabe} presents the master key setup delay and secret key generation delay as a function of the number of attributes. The results demonstrate that the delay almost linearly increases with the number of attributes. The master key setup is independent of the number of attributes, and that of the laptop and Nexus One is given as 166 ms and 354 ms, respectively. Based on our user experience, the key setup delay in the laptop was not noticeable, but that of the mobile device was not ignorable. However, the key setups are not frequent events in PCN and these can be undertaken prior to real time use. Considering the advancements in mobile hardware technology, the key setup delay will be ignorable in the near future.

\emph{Detailed performance of each component}: We measured the performance of the remote content retrieval (single hop). Retrieving/decrypting involves six steps: (1) FUSE open/read call (in laptops only), (2) CCN data retrieval over a remote node, (3) ABE local repository key look-up, (4) ABE public key retrieval from a remote node, (5) CP-ABE decryption of an AES key, and (6) content decryption with AES-256.

As shown in Table \ref{tb:d2dm}, the most time consuming operations are the CCN retrieval and CP-ABE decryption, and they have a positive relationship with the file size. Unlike the CCN retrieval and CP-ABE encryption, the ABE local repository key look-up, ABE private/public key retrieval from a remote node, and CP-ABE decryption of an AES key increase linearly with the file size. The cost of the FUSE operations took less than a few milliseconds, and we did not report a delay in the table. Based on the results, the laptop to Nexus One (L2M) and desktop computer to Nexus One (D2M) exhibited similar behaviors in the CCN retrieval and CP-ABE decryption. However, the desktop computer to laptop (D2L) was 2 $\sim$ 5 times faster than L2M and D2M. These differences were caused by the lower computation power of the Nexus One.

\emph{PCN overhead}: Given that there is no available realistic trace of personal content management, we collected the recently accessed files from Windows PCs. We drew participants from researchers and graduate students, and collected data from a total of 31 participants (25 male, 6 female). The participants varied in age: 22 were between 21-30 years old; 14 were between 31-40 years old; 1 was between 41-50 years old. Windows maintains a link to each file accessed in a designated directory (e.g. Windows XP in the Recent directory). A symbolic link file is automatically created when the target file is opened for the first time; also, whenever the target file is accessed, the link's modification time is automatically updated (refreshed). Although this data set does not provide a real-time trace, it provides valuable information for system evaluations (e.g. the average number of prefixes and content access/update patterns). Our investigation demonstrates that the time span of recently accessed files typically ranges from one to two months. There was only one participant who recently erased their access history, and we excluded this participant from the analyses.

For each user, we plot the number of distinct files and the number of distinct directories (Figure \ref{fig:access}). The number of distinct directories is similar to the number of prefixes that PCN needs to announce (assuming that updates occur therein). The number of distinct files provides a rough usage activity level: users 1 and 2 were less active, whereas users 29 and 30 were more active than the regular users. It appears that most participants accessed 120-140 files over the time span of one to two months; this number is conservative in that it only counts the files that were accessed via file browsers. The figure also demonstrates that the number of directories was typically less than 60. This number is closely related to the number of distinct prefixes announced by the user (which can also be aggregated, e.g. ``\emph{/My Doc/}'').

For a given social network with a hop limit of $k$, the routing table size in an intermediate node is proportional to the number of distinct prefixes. The following is a simple back-of-the-envelope calculation. Assuming that there is an average branching factor of 100 and a hop limit of $k=1$, $k=2$, a node needs to keep the entries of a total of 100 and 10,000 people, respectively. If a user has 60 prefixes, each of which is 100 bytes, the total number of entries is 6,000 and 600,000, respectively, and the total storage demand is 0.6 MB and 60 MB, respectively. The overhead of $k=2$ can be reduced if we selectively include friends of friends because the purpose of extending more than one hop is to limit the impact of Network Address Translation (NAT).

We also analyzed how many files are accessed per directory (Figure \ref{fig:dir}). The figure demonstrates that the number of distinct files accessed per directory is highly skewed, and only a small number of files are accessed per directory. For example, in 41\% of directories, only a single file was accessed, and the fraction of directories that had more than five distinct files accessed was only 10.2\%. This result explains why participants have a high number of directories as opposed to a distinct number of files.

Finally, in order to analyze how people interact with files, we measured the time difference between the link creation and modification (i.e. access time span of a file) and plotted the results in Figure \ref{fig:delay}. Recall that a link is modified (refreshed) whenever the target file is accessed. Interestingly, the figure demonstrates that the file access patterns of personal content is almost \emph{bimodal}; that is, quite a significant percentage of files are only accessed once and are read-only (i.e. time difference is 0, 64.1\%), and another significant percent of files are repeatedly accessed over the time span of longer than one hour (i.e. the time difference is greater than 60 minutes, 28.6\%). The remainder of the files (7.3\%) has intermediate access and are likely be repeatedly accessed over time. The files that are repeatedly accessed include both read-only and read-write accesses. It is expected that the percentage of read-write accesses would be significantly smaller than that of read-only accesses. This also indicates that the overhead of maintaining consistency over personal content networking in practice would be minimal (e.g. only a small number of files are updated over the course of a day).

\begin{table*}[t]\footnotesize
\begin{center}
  \begin{tabular}{ | c | c | c | c | c | c | c | c | c | c |}
\hline
  &  Naming &  DTN &  Topology &   Replication Unit &  Update &   Trust & Access Control &  Secure Binding \\
\hline
\hline Ficus &  SP+H &  Yes &  P2P &   File/Dir &  Yes &   - & ACL &  - \\
\hline BlueFS/EnsemBlue &  SP+H &  Yes &  C/S &   File &  Yes &   - & ACL &  - \\
\hline UIA/Eyo &  DP+H &  Yes &  P2P &   - &  - &   - & ACL &  - \\
\hline PersonalRAID &  SP+H &  Yes &  P2P &   Volume &  Yes &   - & - &  - \\
\hline Footloose &  SP-F &  Yes &  P2P &   File &  Yes &   - & - &  - \\
\hline DisCFS &  SP-H &  No &  C/S &   Volume &  Yes &   KeyNote & Certs &  - \\
\hline Bayou &  SQL &  Yes &  P2P &   Volume &  Yes &   PKI & Certs &  - \\
\hline Plutus/SiRiUS &  SP-H &  No &  C/S &   - &  - &   PKI & Certs/Enc-PKC &  - \\
\hline PAST &  SP-F &  No &  P2P &   File &  Yes &   PKI & Certs/Enc-PKC &  - \\
\hline CCN &  SP-H &  Yes &  P2P &   File &  No &   PKI & Certs/Enc-PKC &  Yes \\
\hline\hline PCN &  SP-H &  Yes &  P2P &  File/Dir &  Yes &   SPKI & Certs/Enc-ABE &  Yes \\
\hline
  \end{tabular}
\caption{Feature comparison: DTN (Delay Tolerant Networking), SP/DP (Single Persistent or Device Persistent), F/H (Flat/Hierarchical), PKC (Public-Key Cryptography)}\inv\inv\inv\sinv
\label{tb:summary}
\end{center}
\end{table*}

\section{Discussion}
\emph{Security attacks}: PCN shares the security benefits of CCN because it is pull-based content retrieval and uses secure binding, thereby effectively thwarting distributed denial of service attacks, request flooding attacks, and man-in-the-middle attacks~\cite{Jacobson09}. While PCN introduces new features such as extra prefix announcements (modification and revocation) and content updates, PCN's explicit prefix protection provides a restriction that only authorized users can replicate a named prefix. Moreover, PCN provides a limitation that replicated content can only be updated by users with explicit write permissions. Thus, a user can neither request content replication nor inject updates without explicit permissions from the content owner, as illegitimate requests are automatically discarded by the intermediate PCN nodes.

\emph{Semantic vs. hierarchical naming}: PCN uses single persistent hierarchical naming. An alternative is semantic naming as in semantic file systems where semantic information is added to file systems and semantic attribute queries are used to locate files~\cite{Gifford91,Geambasu07,Salmon09}. An extreme case would be using a single flat directory where each file has an arbitrary unique name, and a user can search for any files using semantic queries; the user can maintain views across multiple devices~\cite{Salmon09}. However, extensive human subject studies in the personal information management field have demonstrated that a majority of people want to search by browsing a hierarchical file system (called ``orienteering behavior'') and use semantic queries (e.g. desktop search tools) as a last resort~\cite{Jones07}. This results from browsing relying more on recognition and people use browsing to reduce and distribute the amount that must be recalled~\cite{Lansdale88}. Given that only a handful of applications require semantic naming (e.g. music players), it is more efficient to implement semantic data access as an application layer service over PCN.

\emph{Energy efficiency}: The PCN system includes battery powered personal devices. Battery limited devices need to constantly listen to announcement messages, which prevents them from switching to a sleep mode for power saving. Recall that whenever there are updates, PCN broadcasts the messages to the $k$-hop neighbors in the overlay network. One solution to this problem is to introduce a \emph{proxy server} in an AC powered device (e.g. desktop computer, laptop, etc.). A mobile device can re-configure the underlying overlay network topology such that messages always travel through the local proxy server. The local proxy buffers all incoming announcements. Then, the mobile client periodically wakes up and pulls the aggregated announcements.

\emph{Interest-based push for synchronization}: In our prototype implementation, we used prefix announcements to notify replica nodes of content updates. An alternative to this approach is to use \emph{interest solicitation}, as recommended in the NDN proposal~\cite{Zhang10}. The node that updated the content sends an interest solicitation packet to the replica nodes that are interested in receiving the updated content. Then, those interested replica nodes will send an interest packet requesting the updated content. For efficient synchronization, the interest solicitation packet includes detailed information about the updated content because the prefix announcement was augmented in PCN.

\emph{Private PCN}: For security reasons, a user could have two different namespaces: one for private access and the other for shared access. The private PCN is not visible to other users; thus, a user can simplify the access control, e.g. just setting a single attribute for content encryption. Given that a large percentage of content is personal use only, it is expected that a private PCN network could lower the burden of content management.

\emph{Offline devices}: If devices are offline, a user cannot browse the content stored in the devices. In order to aid content retrieval from offline devices, PCN can take a similar approach to that used in Eyo~\cite{Strauss09}. Each device periodically pulls the content lists of the other devices and stores them in its local repository. Given this information, PCN nodes can tell which device has a file and, thus, a user can access the file from offline devices.

\emph{Overlay construction of the devices behind NATs}: A device may be behind a NAT, and it cannot actively participate in the overlay network. In this case, the device can be connected through a relay node that is not behind a NAT and is sufficiently stable (e.g. the device is online 90\% of the time). The NAT can potentially reduce the number of peering devices, thus lowering the connectivity among devices. We can increase the connectivity through allowing devices to exchange IP addresses of $k$-hop friends' devices. For example, when $k=2$, Alice can connect to Bob and also to Bob's friends. Furthermore, computing resources in cloud systems could be utilized to increase connectivity, e.g. a personal account in Dropbox could serve as an intermediate node in PCN.

\section{Conclusion}
We designed and implemented the Personal Content Networking (PCN) platform. We extended the CCN to build a basic framework for distributed content management with replication and updates, and then we implemented a \emph{secure content-centric access control} mechanism using the recently proposed cryptography tool called Attribute-Based Encryption (ABE) that permits selective content sharing over untrusted nodes. The primary departure from prior work is that PCN supports ABE-based secure \emph{read-write} operations over untrusted devices and \emph{secure-binding} between the name and data. We built a PCN prototype through integrating the whole system using a user level file system, and we demonstrated its feasibility through performance measurements and trace analysis.

\section{Acknowledgement}
This work was partly supported by the ICT R\&D program of MSIP/IITP [1391104004, Development of Device Collaborative Giga-Level Smart Cloudlet Technology].

%\begin{acknowledgements}
%\end{acknowledgements}

\bibliographystyle{spmpsci}      % mathematics and physical sciences

\bibliography{CCNFS_WPC}

\end{document}